\documentclass[letter]{aa} 
\usepackage{graphicx}
\usepackage{natbib}

\usepackage{txfonts}
\begin{document}
   \title{Massive star models with magnetic braking}


   \author{Georges Meynet, Patrick Eggenberger
          \and
          Andr\'e Maeder
          }

   \institute{Geneva Observatory, Geneva University, CH-1290 Sauverny, Switzerland\\
              \email{georges.meynet@unige.ch}
              \email{Patrick.Eggenberger@unige.ch}
              \email{Andre.Maeder@unige.ch}
             }

   \date{Received September 15, 1996; accepted March 16, 1997}

 
  \abstract
   {Magnetic fields at the surface of a few early-type stars have been directly detected. These fields have magnitudes between
   a few hundred G up to a few kG. In one case, evidence of magnetic braking has been found.}
   {We investigate the effects of magnetic braking on the evolution of rotating ($\upsilon_{\rm ini}$=200 km s$^{-1}$) 10 M$_\odot$ stellar models at solar metallicity during the main-sequence (MS) phase.}
   {The magnetic braking process is included in our stellar models according to the formalism deduced from 2D MHD simulations of magnetic wind confinement
   by ud-Doula and co-workers. Various
   assumptions are made regarding both the magnitude of the magnetic field  and of the efficiency of the angular momentum transport mechanisms
   in the stellar interior.}
   {When magnetic braking occurs in models with differential rotation,  a strong and rapid mixing is obtained at the surface accompanied
   by a rapid decrease in the surface velocity.  
   Such a process might account for some MS stars showing strong mixing 
   and low surface velocities. 
   When solid-body rotation is imposed in the interior, the star is slowed down so rapidly that
   surface enrichments are smaller than in similar models with no magnetic braking. 
   In both kinds of models (differentially or uniformly rotating), magnetic braking due to a field of a few 100 G  significantly
   reduces the angular momentum of the core during the MS phase. This reduction is much greater in solid-body rotating models.}
   {}


   \maketitle
%

\section{Introduction}

Many indirect observations indicate that magnetic fields should be 
present  around massive stars \citep{Henrichs2005,Hubrig2008}, and recently a few direct detections have been made via the Zeeman effect \citep[see the recent review by][]{Donati2009}.
For instance, direct measurements have been obtained for a handful of O stars, among them  $\theta^1$ Ori C and HD 191612, with values of a
few hundred G \citep{Donati2002, Donati2006a}. This last star is a member of the class of Of?p stars, of which only 5 are known in the Galaxy. Of these, three
have been observed in detail and all three have detectable magnetic fields \citep{Wade2010}.

Among early B-type stars, a magnetic field has been discovered in the B0.2V star $\tau$ Sco \citep[][$\sim$ 0.5 kG]{Donati2006b} and in three
$\beta$ Cepheid stars: $\xi^1$ CMA \citep[B0.7IV, 300 G] {Hubrig2006}, $\beta$ Cep \citep[B2III, less than 100 G]{Henrichs2000}, and V2052 Oph \citep[B2IV, less than 100 G]{Neiner2003b}.
\citet{Neiner2003a} and \citet{Hubrig2006} have discovered magnetic fields of the order of a few hundred G in slowly pulsating B-type stars.
Magnetic fields have been detected around a few Be stars \citep{Yudin2009, Hubrig2007, Neiner2005}, and these
fields are of the order of 100 G or less.

The origin of these magnetic fields is still unknown. It might be fossil fields \citep[e.g. the
spectral characteristics of  Of?p stars are indicative of organized magnetic fields, most likely 
of a fossil origin according to][]{Wade2010}, or fields produced through a dynamo mechanism. Recent simulations by  \citet{Cantiello2010}
of subsurface convective zone in massive stars show dynamo-generated magnetic fields of the order of one kG. According to these authors, these 
magnetic fields might reach the surface of OB stars.

In the Sun, magnetic braking results from solar wind material following the magnetic field lines that extend 
well beyond the stellar surface. This coupling exerts a torque on the surface layers of the Sun, and this slows down its rotation.
Could such a process also be active in massive stars showing a sufficiently strong surface magnetic field?
\citet{Townsend2010} has recently discovered that $\sigma$ Ori E (B2Vpe, $\sim$ 10 kG)  is undergoing rotational braking.
The spin-down time of 1.34 Myr is in good agreement with theoretical predictions based on 
magnetohydrodynamical simulations of angular momentum loss from a magnetized line-driven wind.
This gives some support to the hypothesis that at least a few massive stars may indeed suffer magnetic braking.   

Such an effect has for the moment never been included in massive star models; however,
by modifying the internal distribution of $\Omega$, the angular velocity, magnetic braking can significantly change
the mixing of the elements inside the star, as well as the evolution of its angular momentum content. 

Although rotating models have improved the agreement between models and theory, some
points remain to be clarified.
A small subset
of stars exhibit surface properties, such as low surface velocities and strong surface enrichments, 
which, when the star is not a giant or a supergiant, cannot be explained by current rotating stellar models for single stars
\citep[see discussions in][]{Brott2009, Hunter2009}. 
Although the fraction of these stars is small  and some may be stars at the
end of the MS phase  \citep{Maeder2009a}, it is worthwhile
investigating what  the physical
cause of this behaviour could be. 

Theoretical models with rotation also predict
 rotation velocities  that are too high for the young pulsars \citep[see the discussion in][]{Heger2004, Heger2005}. 
Thus it does appear interesting to study any effects that can remove angular momentum from the star.
Can magnetic braking be an interesting explanation for the strongly mixed, slow rotators? Can it help
in reducing the angular momentum of the core?
In this first paper, we want to study these questions by presenting a first series of computations accounting for this effect.
In section 2 we present the formalism we used to account for the magnetic braking effect,
as well as the physical ingredients of the models. The results are discussed in sections 3 and 4 and conclusions are given in section 5.




\section{Physics of the stellar models}

  \begin{figure}[t]
   \centering
   \includegraphics[width=8cm]{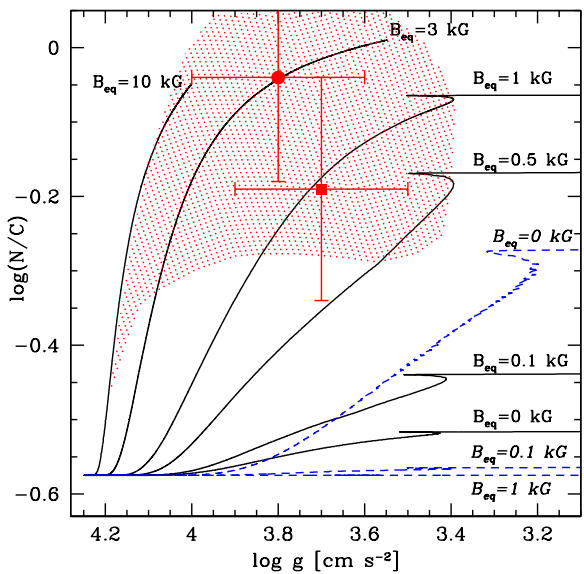}
      \caption{Evolution of the N/C ratio (in number) as a function of the gravity at the surface of 10 M$_\odot$ stellar models 
      with $\upsilon_{\rm ini}=200$ km s$^{-1}$. The value of 
      the magnetic field is indicated.
      Continuous (dashed) lines are for models with interior differential (solid body) rotation. In the shaded areas, models with differential rotation predict surface velocities
      inferior to 50 km s$^{-1}$ during the MS phase. The full circle and square
      correspond to HD 16582 and HD 3360 ($\zeta$ Cas) respectively (see text for references). 
              }
         \label{NC}
   \end{figure}
   
     \begin{figure}[t]
   \centering
   \includegraphics[width=8cm]{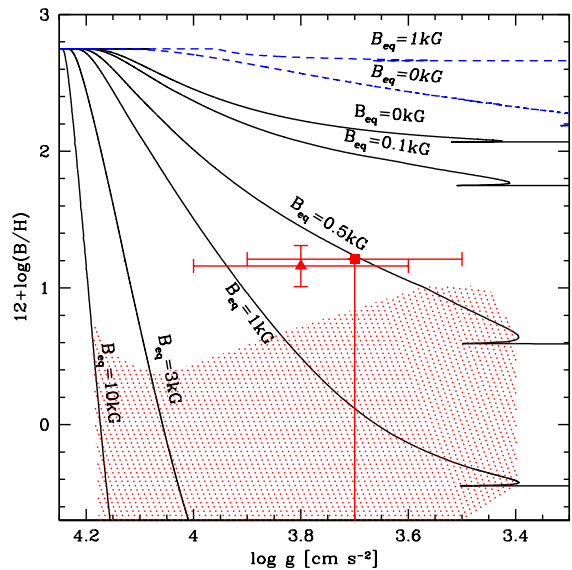}
      \caption{Same as Fig.~\ref{NC} for the evolution of the abundance of B at the surface.
              }
         \label{bore}
   \end{figure}

The formalism of the magnetic braking law used here follows theoretical developments
first made for the Sun, starting with
\citet{Weber1967} who used an idealized monopole field to model the angular momentum loss in the solar wind $\dot J$.
They found that 
$
\dot J={2/3} \dot M \Omega R^2_{\rm A},
\label{WD}
$
where $\dot M$ is the mass loss rate and $R_{\rm A}$ the Alfv\'en radius, defined as the point where the ratio between the magnetic field energy density and the kinetic energy density of the wind is equal to 1.
\citet{UdDoula2002} and \citet{UdDoula2008} have examined the angular momentum loss from magnetic hot stars with a line-driven stellar wind and a rotation-aligned dipole magnetic field using 2-D numerical MHD simulations.
They find that the total angular momentum loss 
follows
the expression given above, but with a smaller $R_{\rm A}$ than in the monopole case.
They find that
\begin{equation}
{{\rm d}J\over {\rm d}t }={2 \over 3}\dot M \Omega R^2_*[0.29+(\eta_* + 0.25)^{1/4}]^2,
\label{ma}
\end{equation}
where $\eta_*$ is the magnetic confinement parameter \citep{UdDoula2002} defined by 
$
\eta_* \equiv {B^2_{\rm eq} R^2_* / \dot M \upsilon_\infty},
\label{eta}
$
where $R_*$ is the stellar radius, $B_{\rm eq}$, the magnetic field at the equator and at the surface of the star, and $\upsilon_{\infty}$ the wind velocity at infinity.
Very interestingly, when applied to the case of $\sigma$ Ori E, Eq.~\ref{ma} gives a  spin-down timescale of the order of 1 Myr \citep{UdDoula2009}, quite
in agreement with the observed spin-down timescale for this star \citep{Townsend2010}. 

We have implemented in the Geneva evolution code \citep{Eggenberger2008} Eq.~\ref{ma} for the loss of the angular momentum.
Various values of $B_{\rm eq}$ are used between 0.1 and 10 kG. In the present exploratory work, we suppose that the magnetic field keeps a constant
value during the MS phase.
We computed the evolution of 10 M$_\odot$ stellar models at solar metallicity with $\upsilon_{\rm ini}$= 200 km s$^{-1}$.
In the context of shellular rotation \citep{Zahn1992}, the vertical transport of chemicals
through the combined action of vertical advection and strong
horizontal diffusion 
can be described as a pure diffusive process \citep{Chaboyer1992}.
The advective transport is then replaced by a diffusive term, 
with an effective
diffusion coefficient $D_{\rm eff} ={|rU(r)|^2}/{30D_{\rm h}}$,
where $r$ is the characteristic radius of the isobar, $U(r)$ the vertical component of the meridional circulation velocity \citep{Maeder1998} and $D_{\rm h}$ the diffusion coefficient associated to horizontal turbulence as given by \cite{Zahn1992}. The vertical transport of chemical elements then
obeys a diffusion equation, which in addition to this macroscopic transport,
also accounts for (vertical) turbulent transport with the same coefficient 
$D_{\rm shear}$ as for the transport of angular momentum \citep{Maeder2001}.

The effects of magnetic braking depend a lot on how angular momentum, hence the chemical species, is
transported inside the star. Here we consider two limiting cases:
\begin{itemize}
\item {\bf Differential rotation:} the angular momentum transport is driven by the meridional currents and the shear instabilities. These two effects
are not efficient enough to produce solid body rotation. A moderate differential rotation is therefore present during the whole MS 
phase, and the mixing of the chemical species is mainly due to shear instabilities.
\item {\bf Solid body rotation:} when the transport of the angular momentum is very efficient, then 
solid body rotation is maintained during the whole MS phase\footnote{We do not invoke any
specific mechanism here. Solid-body rotation might be due to a fossil magnetic field for instance or to a dynamo driven by shear such as the one proposed by \citet{Spruit2002}.}.  The chemical species are transported by the meridional currents. 
\end{itemize}
At the present time it is not known which is the most frequent situation in nature, so
we explore these two possibilities, which represent some limiting cases hopefully framing the real world.
 
\section{Models with magnetic braking and interior differential rotation}

  \begin{figure}[t]
   \centering
   \includegraphics[width=8cm]{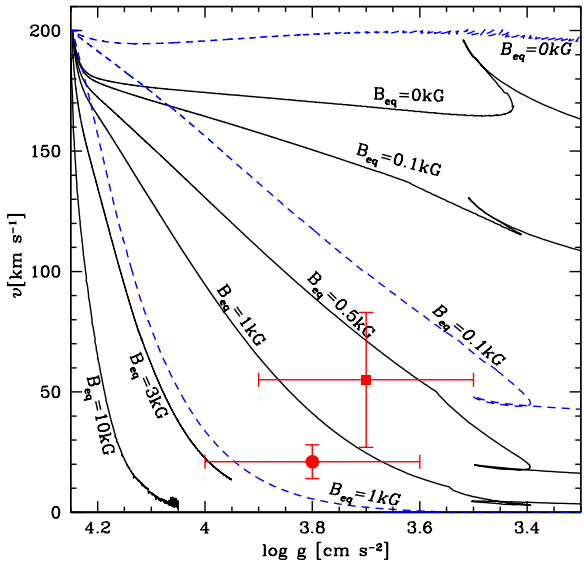}
      \caption{Same as Fig.~\ref{NC} for the evolution of the surface velocity at the equator.
              }
         \label{Veq}
   \end{figure}
   
   In Figs.~\ref{NC}, \ref{bore},  and \ref{Veq} (look at the continuous lines) are shown the evolutions of the abundances and of the (equatorial) velocities at the surface
   of models computed with different values of the magnetic field. We see that, compared with the model without magnetic braking, models with 
   magnetic braking present stronger changes of the surface abundances.
   This comes from the fact that magnetic braking creates strong differential rotation in the outer layers and thus triggers mixing by shear. At the same time
   the surface velocity rapidly decreases. 
   
   More quantitatively, we see that, at 100 G, the effects on N/C remain modest
   (increase by about 20\% at the end of the MS phase with respect to the model without magnetic braking) but are already significant for the surface velocity, which is decreased from 168 to 115 km s$^{-1}$
   at the blue turnoff of the MS band. For a 500~G magnetic field, the effects are very strong: increase by more than a factor 2.2 of the N/C ratio at the end of the MS phase. The surface velocity
   reaches values around 19 km s$^{-1}$ at the end of the MS phase. 
   Surface abundances of boron provide a very sensitive test of how mixing occurs in the outer layers of the star. Indeed, that element already begins to be destroyed
   at  temperatures between 5-6 10$^6$ K. From Fig.~\ref{bore}, we see that in models with magnetic braking, boron is much more rapidly depleted at the surface. 
   
   Stars in the hatched zone of Figs.~\ref{NC} and  \ref{bore} present 
   strong signs of mixing during the main sequence and have  surface velocities inferior to 50 km s$^{-1}$. Typically, one would expect that at least some
   LMC stars in group 2 of Fig. 5 in \citet{Hunter2009} would be in this region. We defer a detailed comparison with the data of \citet{Hunter2009} to a later paper
   where models with the LMC metallicity will be computed.
    In Figs.~\ref{NC}, \ref{bore},  and \ref{Veq}, we have indicated
   the positions of the two known stars 
   having initial masses between 9 and 11 M$_\odot$ and showing boron depletion by more than an order of magnitude \citep[data taken from 
   Tables 3 and 4 of ][see detailed references therein]{Frischknecht2010}. 
   We see that current models with magnetic braking although not providing a perfectly
   consistent fit to the data\footnote{All our models start their evolution with a velocity
   of 200 km s$^{-1}$, which may not be adequate for these stars.}, appear to be promising for explaining stars with strong signs of mixing,  low surface velocities, and
   high gravities. It is interesting to mention
   that the star corresponding to the square ($\zeta$ Cas) has a detected magnetic polar field of 335($+$120$-$65) G \citep{Neiner2003a}.
   
   Can this process extract angular momentum from the core region? 
 In Fig.~\ref{jspec}, the specific angular momentum inside the central 2 M$_\odot$ at the end of the MS phase
   is shown for various models. 
   We recall that the reduction between the angular momentum of the core on the ZAMS and at the presupernova stage occurs for a significant part
   during the MS phase. Typically about 70\% of the total angular momentum lost by the core during the whole stellar lifetime is lost
   during the MS phase \citep[see Fig. 7 in][]{Hirschi2004a}.
   In models with differential rotation, a magnetic field of about 100 G reduces the specific angular momentum by about 25\% (0.1 dex) with respect
   to the values obtained in the model without magnetic braking.  When a magnetic field between 500 G and 1 kG  is considered, $j$ is on average decreased by slightly more that 0.3 dex, which means
   by a factor between 2 and 2.2 with respect to the model with no magnetic braking.  
   Thus magnetic braking in differentially rotating models has an impact on the evolution of core angular momentum content. 
   
   
    \section{Models with magnetic braking and interior solid body rotation}

  \begin{figure}[t]
   \centering
   \includegraphics[width=8cm]{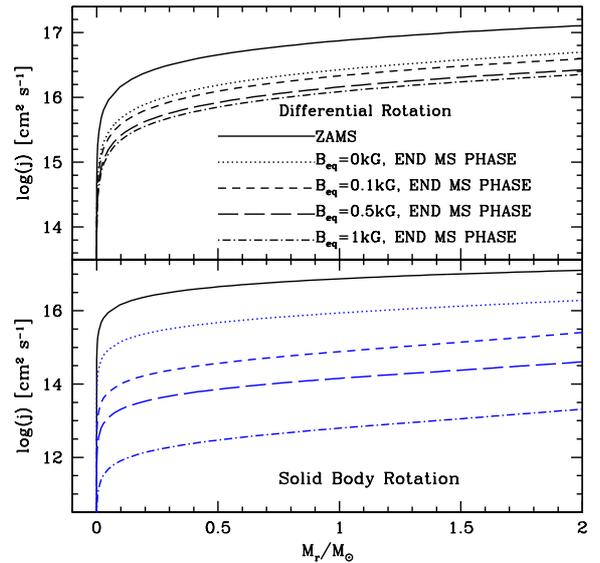}
      \caption{Variation in the specific angular momentum as a function of the Lagrangian mass inside the cores of different models.}
         \label{jspec}
   \end{figure}
   

The case of solid body rotation is quite different from the one of differential rotation as can be seen in Figs.~\ref{NC}, \ref{bore}, \ref{Veq}, and \ref{jspec}. This is expected since the physical
processes responsible for the transport mechanisms are quite different. For the angular momentum, a strong coupling is imposed in the model. The chemical elements are no longer
transported by shear turbulence as in the previous case, but by meridional currents. 

Whenever meridional currents are responsible for the transport of the chemical species, the key factor governing chemical element mixing is $\Omega$, which decreases rapidly in the whole
interior when magnetic braking is exerted at the surface.  When magnetic braking is accounted for, mixing is much weaker than in the case without magnetic braking. For a magnetic field
equal to or larger than 0.5 kG, no nitrogen enrichment is predicted by the models during the whole MS phase. Boron depletions
are not as strong in models with magnetic braking.

From Fig.~\ref{Veq}, we can see that for models with magnetic braking and solid body rotation, 
the surface rotation decreases much more rapidly than in the
case of differential rotation (for a given value of $B_{\rm eq}$).  Thus such models would predict
stars at the end of the MS phase with low surface velocity and weak or even nonexistent surface enrichments.
Such stars can also result from the evolution of progenitors starting with low initial rotational velocities.
However, the internal structure at the end of the MS will be different if the star has ever been a slow rotator or
has come from a slowed down, initially rapid rotators. Typically the star with a fast-rotating progenitor will have
a larger core. 


Solid body rotation alone (without magnetic braking) is already an interesting way to evacuate the angular momentum enclosed in the core.
For instance, it has been shown by \citet{Heger2005, Heger2004}, who computed models where solid body rotation is imposed during the MS phase by the action of the Tayler-Spruit dynamo
\citep{Spruit2002}, that these models
could produce neutron stars with rotational velocities at birth in the upper range values given by observations. 
As can be seen from Fig.~\ref{jspec} (lower panel),
magnetic braking with solid body rotation has a dramatic effect on the angular momentum content of the core.
We see that, already with a value of $B_{\rm eq}$ = 100 G, $j$ is decreased by more than one dex 
with respect
to the corresponding solid-body rotation with no magnetic braking. For magnetic fields of 1 kG, $j$ is decreased by more than
4 orders of magnitude! In that last model, the specific angular momentum in the core at the end of the MS phase has the same order of magnitude as
the one required for explaining the long-period (150 ms) young pulsar PSR B1509-58 \citep[see Table 3 in][]{Heger2004}.
%

\section{Conclusions}

The main results in this paper are the following.
Models with differential rotation and magnetic braking may produce strongly mixed stars with low
surface velocities during the MS phase. 
Some of the stars discussed by \citet[][ see their Group 2 stars in their Fig. 5]{Hunter2009}
as well as some stars showing very strong boron depletion with no N-enhancement \citep{Venn2002, Mendel2006, Frischknecht2010},
might be stars that suffered magnetic braking. Models with solid-body rotation and magnetic braking produce stars that at the end of the MS phase have
low surface velocities and none or very weak changes in the surface abundances. 
Magnetic braking
reduces the angular momentum content of the core, even for moderate values
of the magnetic field when solid body rotation is considered. 

Present models are of course quite preliminary and have to be improved in the future.
For instance,  the coexistence of a large-scale/dipolar magnetic field (as adopted here) 
with internal differential rotation may be problematic. If the field is of fossil origin, it will likely pervade the whole star
and may induce strong coupling, thus reducing or even removing any differential rotation. In that case,
our models with solid-body rotation would apply. In these models, magnetic braking is so efficient that
the slow down occurs before any significant changes in the surface abundances.
If the surface magnetic fields are produced by a dynamo \citep[see for instance the suggestion by][]{Cantiello2010},
differential rotation may still be present in the interior, and changes in the surface abundances
are quite rapid. However,
one may question whether it is justified to assume a constant value for $B_{\rm eq}$ as we did here.
More quantitative estimates need detailed understanding of the dynamo 
process and of its dependence on rotation. 
Finally let us stress  that in the Hunter diagram most stars follow the predictions
of rotating models  \citep{Maeder2009a} without magnetic braking. This is consistent with the weak
magnetic fields of OB stars.

\bibliographystyle{aa}
\bibliography{MyBiblio}

\end{document}